\documentclass{article}
\usepackage{graphicx} 

\title{Ideal Social Gas: Emergent Thermodynamic Observables in an Effective Model of Social Dynamics}
\author{V.S. Morales-Salgado}
\date{}

\begin{document}

\maketitle

\begin{abstract}
 This work continues the one commenced in \cite{m24}, where the key idea is that individual stances on a social matter can be modeled as positions of particles in physics.
 Here, we develop an effective thermodynamic framework for the statistical description of collective social systems based on a mechanical representation of individual stances. In this approach, each individual is modeled as a point-like particle evolving in an abstract stance-space, where the dynamics are characterized by position-dependent inertial response functions. 
 The resulting many-particle system is treated within an equilibrium statistical description analogous to the canonical ensemble of statistical mechanics.
 
 Using this framework, we derive the corresponding partition function and introduce emergent macroscopic observables analogous to pressure, volume and temperature, interpreted here as collective statistical properties of the social system. 
 For a particular class of position-dependent mass functions, the resulting equation of state acquires an ideal-gas-like form, suggesting a degree of macroscopic universality despite microscopic heterogeneity among individuals.
 The formalism is further extended to open social systems through the introduction of a chemical potential associated with population exchange among social groups. 
 This extension naturally allows the treatment of variable particle number and multispecies social configurations within the same equilibrium framework.
 
 The present work is intended as an exploratory phenomenological contribution to sociophysics and complex systems research, aimed at investigating whether collective social behavior may admit effective macroscopic statistical descriptions analogous to those encountered in statistical physics.
\end{abstract}

\section{Introduction}
This exploration belongs to the discipline known as \emph{social physics} \cite{j22,w69}, where concepts and tools from physics are used to study social phenomena.
Here we are interested in constructing an effective description of members
of a social group in terms of point-like particles, where the position of
each particle represents the stance (opinion, belief, conviction) of the
corresponding individual.
Hence, this work explores the possibility of using a physical system to investigate a social system through a so-called \emph{analogue model} \cite{s17}.

Although a social phenomenon is realized by a group of individuals, it shows richer properties than just the aggregation of individual behavior \cite{ak14}.
In this sense, there is an emergence of social phenomena that naturally divides the study of human behavior into two realms: the individual and the social behaviors \cite{g04}.

This seemingly paradoxical observation is also encountered in physics, where there is a natural divide between the microscopic and macroscopic scales.
One can find a particular example in how thermodynamical phenomena emerge from a mechanical description of particles, along with some statistical considerations in what is called \emph{statistical mechanics}.

Hence, a statistical treatment may provide a useful phenomenological framework for describing collective social behavior through emergent
macroscopic variables.
That is precisely the purpose of this work.
However, note that this analogy was already proposed in \cite{p96}.
This, in turn, requires a description of individual behavior analogous to the mechanics of particles.
This has already been done in \cite{m24}.
A brief recount of the results is presented in Section \ref{Particles} of this work.

Departing from that previous work, here we are interested in making some assumptions that allow us to obtain an initial model that, even if simple, allows us to describe general features of social systems.
This reasoning is presented in Section \ref{Gases}, where the partition function is introduced along with the workhorse model of the rest of this article: particles with mass functions that are quadratic with respect to its position.

The resulting statistical description allows the introduction of collective
observables characterizing the macroscopic behavior of the social system.
To achieve this, the equation of state in terms of the partition function is introduced in Section \ref{Thermodynamics}.
This is motivated by reinterpreting the thermodynamical variables that describe gases now in the context of social phenomena.

In Section \ref{Discussion}, we reflect on the consequences of the analogous model presented here.
This helps us to further explain how the results coherently describe social phenomena in terms of relations among macroscopic (social) variables.
Finally, the concluding remarks and some prospective work are given in Section \ref{Conclusions}.

It is important to emphasize that the purpose of the present work is not to
construct a realistic many-body theory of social systems in full generality, but rather to investigate whether a coherent macroscopic description can emerge from simple microscopic assumptions, analogously to the historical role played by the ideal gas model in statistical physics. 
Consequently, the assumptions introduced here, including equilibrium and non-interacting particles, should be understood as effective idealizations that allow us to explore the possibility of defining collective social variables through statistical methods.

\section{Particle mechanics}\label{Particles}
 In \cite{m24} we introduced a phenomenological mechanical framework in which individual stances are represented by effective positions in an abstract configuration space.
 The analogy relies on the ideal category of particles, so that an opinion is represented by a particle and their evolution by a trajectory.
 Other required analogies such as inertia, mass, momentum, interaction, and force were also introduced and explained.
 
 The main result of that work was the plausibility of having an equivalent of \emph{Newton's second law} in mechanics, where a force $F(x,t)$, which is generally a function of time $t$ and the position of the particle $x$, drives the motion of the particle with a position-dependent mass $m(x)$ through the following law of motion:
    \begin{equation}\label{EoM}
     m(x)\,\ddot{x} + m'(x)\,\dot{x}^2 = F(x,t) \,,
    \end{equation}
 where we use the notation $\dot{f}=\frac{\rm d}{{\rm d}t}f$ and $f'=\frac{\rm d}{{\rm d}x}f$ for an arbitrary function $f$.
 
 Initial concepts and their relations are rather simple in the sense that they pertained to immediate calculations and consequences of representing opinions by point-like objects: particles.
 However, we are warned that other more complicated principles from physics cannot be taken for granted.
 In any case, they should be either derived or strongly motivated in order to include them as an assumption.
 One of these notions is that of \emph{energy}.
 
 For mechanical systems, the concept of energy $E$ is very useful due to its conservation.
 It is commonly defined as composed of two parts: \emph{kinetic energy} defined as:
    \begin{equation}
     K = \frac{m}{2}\dot{x}^2 \,,
    \end{equation}
 and, when one has a conservative force $F=-U'(x)$, the \emph{potential energy} $U(x)$.
 However, there is no reason for \emph{social forces} to be conservative.
 In addition, the position-dependent mass implies that the systems of interest are non-conservative in the sense that, even if we can define $E=K+U$, the system can present losses or gains of the energy thus defined.
 
 Although peculiar, the lack of conservation of energy does not prevent us from continuing to explore the possibilities of translating mathematical methods from physics to the study of social phenomena.
 In the next section, we study an extension of this framework towards a statistical description based on the interaction of many social particles.
 Note that statistical mechanics of physical systems with position-dependent mass has been investigated in \cite{gaat25}.

\section{Social statistical mechanics}\label{Gases}
In this and the next section, we shall follow standard presentations of the ideas, assumptions and results of statistical mechanics, as well as the derivation of the ideal gas law produced from them \cite{pb94}.

Let us suppose a system of $N$ individuals that can be characterized each by the value of their energy $E_i$, $i=1,\dots,N$.
Then let us suppose that the probability to find a particle $i$ in a state of energy $E_i$ that is proportional to the factor:
    \begin{equation}\label{Bfactor}
     {\rm e}^{-\beta E_i} \,,
    \end{equation}
where $\beta$ is a parameter that, for now, particularizes the system as a whole by specifying this distribution with respect to the energy of the individuals that make up it.
At this stage, the Boltzmann factor is introduced as an equilibrium ansatz
motivated by the mathematical structure of statistical mechanics, rather than
as a consequence derived from microscopic social interactions. 
Thus, the distribution in eq.~(\ref{Bfactor}) should be regarded as an effective statistical assumption allowing the exploration of possible macroscopic descriptions of social systems.

In macroscopic equilibrium, we describe the system in terms of the quantity:
    \begin{equation}
     Z = \frac{1}{N!}\int\cdots\int \prod_{i=1}^N {\rm e}^{-\beta E_i}\,{\rm d}E_1\dots{\rm d}E_N \,,
    \end{equation}
known as \emph{partition function}.

In the simplest case of a gas, where all energy is kinetic: $E=\frac{p^2}{2m(x)}$, we have that:
    \begin{equation}
     Z = \frac{1}{N!}\int\cdots\int {\rm e}^{-\beta \sum_i \frac{p_i^2}{2m_i}} \,{\rm d}p_1\dots{\rm d}p_N\,{\rm d}x\dots{\rm d}x_N \,,
    \end{equation}
where we recall that, for our social particles, the mass is a function of the position: $m_i=m_i(x_i)$.
The integral with respect to a single momentum $p_i$, is the so-called \emph{Gaussian integral} that can be readily solved to produce the partition function:
    \begin{equation}
     Z = \frac{1}{N!}\int\cdots\int \prod_{i=1}^N \sqrt{\frac{2\pi m_i(x_i)}{\beta}} \,{\rm d}x \dots{\rm d}x_N \,.
    \end{equation}

To proceed, we need to specify the distribution of the mass function $m_i(x_i)$ of the particles composing the gas: the social inertia of the population in the society of interest.
The case where $m_i$ is the same constant for every particle is the usual monoatomic gas in physics.
Having a set of different constant values $m_i$ for different groups yields a polyatomic gas.
However, for \emph{social matter}, we have the functions $m_i(x_i)$ at our disposal to try to model a society.

In line with having mass characterizing an effective predisposition of each particle toward a preferred position, $m_i(x_i)$ is a distribution of that tendency among the particles labeled by $i$.
For simplicity, let us assume that each of these functions has the same functional form.

The position dependence of the mass function plays a central role in the present framework. 
Unlike ordinary gases in physics, where inertia is an intrinsic constant property of each particle, here the effective inertial response depends on the current stance of the individual. 
Consequently, the distribution of mass functions encodes heterogeneous susceptibilities to social change across the population. 
In this sense, the resulting statistical system is not merely a direct transcription of the physical ideal gas, but a generalized ensemble whose macroscopic behavior depends explicitly on the distribution of social inertial responses.

Let us gain intuition by considering the arguably simple model $m_i(x_i)=a_i\,(x_i+b_i)^2$, where $a_i$, $b_i$ are parameters that characterize each particle.
Then:
    \begin{equation}
     Z =  \frac{1}{N!}\left(\frac{2\pi}{\beta}\right)^{N/2} \int\cdots\int \prod_{i=1}^N \sqrt{a_i}\,(x_i+b_i) \,{\rm d}x \dots{\rm d}x_N \,. \label{PARTN}
    \end{equation}
Also, we need to specify the limits of integration, that is, the allowed stances in the society.
The extent of these limits is analogous to the volume in physical systems.
Consequently, suppose that the particles can be in positions between $-L/2$ and $L/2$.
We thus have:
    \begin{equation}
     Z = \frac{1}{N!}\left(\frac{2\pi}{\beta}\right)^{N/2}L^N \prod_{i=1}^N \sqrt{a_i}\,b_i \,.
    \end{equation}
We can define $G(L,N,a_i,b_i)=L^N\prod_{i=1}^N \sqrt{a_i}\,b_i$ as a quantity that characterizes the total population of particles to obtain:
    \begin{equation}\label{sqgas}
     Z = \frac{1}{N!}\left(\frac{2\pi}{\beta}\right)^{N/2}G(L,N,a_i,b_i) \,.
    \end{equation}
   
\section{Social thermodynamics}\label{Thermodynamics}
The equilibrium description considered here should be understood as an
effective approximation appropriate for sufficiently stable macroscopic social configurations. 
Although realistic societies are generally open, driven and nonequilibrium systems, equilibrium statistical methods provide a natural first step toward investigating whether collective social variables can be defined in a mathematically consistent manner.

Then, now that we have a partition function $Z$, we can use it to obtain some macroscopic properties analogously to the case of physical systems.
The immediate quantities to consider are volume, pressure, and temperature. 
These quantities should be understood as coarse-grained variables emerging
from the statistical aggregation of individual dynamics, rather than as
fundamental properties attached to isolated individuals.
They are understood as follows:
\begin{itemize}
    \item The volume $V$ describes the total region allowed for the particles in the social gas.
    It is limited by boundaries to the displacement of positions (social stances).
    One can imagine the walls of these boundaries to be set by social norms, be they legal, ethical, or otherwise.
    Also, note that the walls of the container need not be rigid.
    We could have a scenario similar to that of a balloon, where the volume can change in time.
    \item The pressure $P$ describes the collective statistical tendency of the population to push against the boundaries of the socially accessible region.
    In analogy with physical gases, pressure emerges from the accumulation of individual motions reaching the limits of allowed stances. 
    Hence, pressure is not associated with a single individual, but rather with the macroscopic effect produced by the aggregated dynamics of the population near the boundaries set by social norms.
    \item The temperature $T$ is a macroscopic measure of the speed with which the particles in the gas move.
    Since this gas is a society, temperature is an overall property of the society being represented that characterizes how quickly the individuals are changing their stances on the topic under consideration. 
    In particular, it should be stressed that the temperature introduced here does not correspond to a physical thermal quantity, but rather to an effective macroscopic parameter characterizing the typical rate of change of individual stances within the population.
\end{itemize}

This analogous interpretation of physical concepts in a social setting shows the prospective usefulness of treating social phenomena analogously to thermodynamical systems.
In particular, we would like to focus on the advantage of being able to characterize a social system in terms of very few parameters, just as thermodynamical parameters broadly describe a many-particle system.

According to this motivation to do thermodynamics on social systems, consider the \emph{free energy} of the system, defined as follows:
    \begin{equation}
     F = -\frac{1}{\beta}\,{\rm ln}Z \,,
    \end{equation}
where $\beta=\frac{1}{k\,T}$ for a proportionality constant $k$.
Although in physics $k$ is a universal constant known as Boltzmann's constant $k_B$, in this analogous treatment of social systems such a statement is rather strong.
Thus, we shall assume that it is different from system to system, although constant for a given one.

In any case, the free energy allows us to obtain the \emph{equation of state} of the social system by defining the pressure of the system as:
    \begin{equation}
     P = -\left(\frac{\partial F}{\partial V}\right)_{\beta,N} \,, \label{EOS}
    \end{equation}
where $F$ depends on $V$ through the fact that the limits of integration in eq. (\ref{PARTN}) that defines $Z$ are given by the boundaries of the region of allowed stances, whose size is the volume.

For the example in the previous section leading to equation (\ref{sqgas}), we have that:
    \begin{equation}
     F = \frac{N}{2\,\beta}\,{\rm ln}\left(\frac{2\pi}{\beta}\right)
        -\frac{1}{\beta}\,{\rm ln}\left(N!\right)
        + \frac{1}{\beta}\sum_{i=1}^N\,{\rm ln}\left[ g(V,\alpha_{i,1},...,\alpha_{1,k})\right] \,.
    \end{equation}
Here, Stirling's approximation, ${\rm ln}(N!)\approx N{\rm ln}N-N$, usually follows. 
Thus, the free energy may be written as follows:
    \begin{equation}
     F = \frac{N}{\beta}\,\left[\frac{1}{2}\,{\rm ln}\left(\frac{2\pi}{\beta}\right)
        -{\rm ln}N + 1\right]
        + \frac{1}{\beta}\sum_{i=1}^N\,{\rm ln}\left[ g(V,\alpha_{i,1},...,\alpha_{1,k})\right] \,.
    \end{equation}
This, in turn, yields the following equation of state:
    \begin{equation}
     \beta\,P = \sum_{i=1}^N\,\frac{1}{g(V,\alpha_{i,1},...,\alpha_{1,k})}\,\frac{\partial}{\partial V}g(V,\alpha_{i,1},...,\alpha_{1,k}) \,.
    \end{equation}
In the simple case where we have no interactions among particles with the same kind of mass function $m(x,\alpha_{i,1},...,\alpha_{1,k})$, we have the following:
    \begin{equation}
     g(L,\alpha_{i,1},...,\alpha_{1,k})=\int_{V/2}^{V/2} \sqrt{m(x,\alpha_{i,1},...,\alpha_{1,k})}{\rm d}x  \,.
    \end{equation}

Once again, if the mass of each particle is given by $m_i(x_i)=a_i\,(x_i+b_i)^2$, then:
    \begin{equation}
    F = \frac{N}{\beta}(\ln N-1) -\frac{N}{2\beta}\ln\left(\frac{2\pi}{\beta}\right) -\frac{1}{2\beta}\sum_{i=1}^N\ln a_i -\frac{1}{\beta}\sum_{i=1}^N\ln b_i -\frac{N}{\beta}\ln L.
    \end{equation}
corresponding to the following equation of state:
     \begin{equation}
     \frac{PV}{T} = Nk \,, \label{SEOS}
    \end{equation}
with $\beta=\frac{1}{k\,T}$.

Thus, for the particular mass function considered here, the resulting
equation of state retains the same functional form as the ideal gas law.
The dependence on the parameters $a_i$ and $b_i$ contributes only as an
overall multiplicative factor in the partition function and therefore does
not explicitly modify the macroscopic equation of state. 

Indeed, this resembles the universality of the ordinary ideal gas equation, whose macroscopic form is likewise independent of several microscopic details of the constituent particles.
More general position-dependent mass functions may produce nontrivial
modifications of the equation of state. For example, a quadratic form such
as
    \begin{equation}
    m(x)=a(1+\lambda x^2)
    \end{equation}
leads to configuration-space integrals with nonlinear dependence on the
volume and therefore modifies the ideal-gas relation.

Note that eq. (\ref{SEOS}) indeed describes the state of the social system. 
The quantities involved are meaningful socially in that they pertain to the macroscopic realm, i.e., they are a result of the aggregated behavior of the total population. 
Therefore, we can say that eq. (\ref{SEOS}) provides us with a thermodynamical description of the social phenomena under study.

\subsection{Chemical potential and variable particle number}
Up to this point, the number of particles $N$ has been assumed fixed.
However, realistic social systems are generally open systems in which individuals may enter or leave a given social group, migrate between communities, or change social affiliation. 
Consequently, it is natural to consider an extension of the present framework in which the particle number is allowed to vary.

In statistical physics, systems with variable particle number are described through the grand canonical ensemble, where the chemical potential $\mu$ appears as the thermodynamic quantity conjugate to $N$. 
Analogously, one may introduce an effective social chemical potential characterizing the exchange of individuals among social groups or stance populations.

The corresponding grand partition function is given by
    \begin{equation}
    \mathcal{Z} = \sum_{N=0}^{\infty} e^{\beta \mu N} Z_N,
    \end{equation}
where $Z_N$ denotes the canonical partition function for a system with $N$
particles.
In this description, equilibrium states are characterized by fixed values of
temperature, volume and chemical potential.

In the present context, the chemical potential may be interpreted as an effective macroscopic parameter describing the statistical tendency of a social system to gain or lose individuals. 
More precisely, $\mu$ quantifies the variation of the effective social energy associated with changes in the population of a given social group or stance configuration.

In this sense, the chemical potential plays a role analogous to a social membership potential: systems with larger $\mu$ tend to attract or retain
individuals more efficiently, while systems with smaller $\mu$ tend to lose population.

Furthermore, different social groups, ideological sectors, or communities may be treated as distinct social species, each characterized by its own chemical potential $\mu_i$. 
The resulting framework allows the description of particle exchange among multiple interacting populations in equilibrium, analogously to multicomponent systems in statistical mechanics.

The corresponding analogue of the thermodynamic differential for the
effective internal energy is then:
    \begin{equation}
    dE = T\,dS - P\,dV + \mu\, dN.
    \end{equation}
Thus, the chemical potential represents the change in effective social energy associated with variations in the number of individuals within the system.

As with the other thermodynamic quantities introduced throughout this work, the chemical potential should not be interpreted as a literal psychological or sociological quantity assigned to individuals. 
Rather, it constitutes an effective collective observable emerging from the statistical description of the population as a whole.

\section{Interpretation of the results}\label{Discussion}
Let us now explore the interpretation of these emergent macroscopic observables within the effective thermodynamic description introduced here.
In terms of positions (stances on a social question), volume represents accessible positions, pressure represents the collective statistical tendency to modify the boundaries of accessible positions.

Some simple relations among volume, pressure, and temperature can be obtained in analogy with physics.
One can look for processes where one of the three variables is kept constant and study the relation between the remaining pair.
Although a full use of the model would of course consider variations of
multiple or all variables, depending on the phenomena at hand, let us
describe the three limiting processes in the context of social mechanics.

Consider first the case of a fixed volume.
The first idealization is where the set of available positions remains fixed.
This constant volume scenario requires fixed walls that artificially do not allow positions to go further than a certain point.
This is the case of well established norms that are not challenged by individuals.
As an effective explanation, the reason behind the rigidity of the wall is not relevant for now.

In physics, higher temperatures yield higher pressures.
This can also be appreciated in the equation (\ref{SEOS}) for the social model presented here, and the reason is statistical:
higher $T$ (lower $\beta$) means faster changes in individual stances that, in turn, imply a larger probability for individuals to reach the boundary of what is socially allowed by norms, described here by the boundary and quantified by $V$.

From this perspective, pressure can be understood as an emergent boundary
effect generated statistically by the motion of the population in stance-space.
In this sense, pressure may be interpreted as a form of social boundary
pressure generated collectively by the dynamics of the population.
Thus, the equation of state does not merely relate abstract variables, but
encodes how the statistical aggregation of microscopic stance dynamics can
produce macroscopic tendencies toward the expansion or preservation of
socially accessible regions.

If those in control of the boundaries wish to maintain them (or even moving them inward, that is, making norms more restrictive), a type of Newton's third law could naturally be proposed: the integrity of the boundary is preserved by exerting a force back on the colliding particle (enforcement of the norm), and the greater the particle's attempt, the greater the reacting force.
In fact, extreme scenarios where the boundaries break are possible in both physical and social contexts.

Alternatively, if the pressure is fixed, we can also appreciate a direct relation between how fast the stances change, indicated by temperature, and the possible stances allowed by the relevant norms, indicated by volume.
A fixed social pressure can be understood as a set of norms enforced with constant intensity. 
This means that even when the population is colliding with the boundaries (set by the norms) with increasing energy and frequency, those in charge of enforcing the norms do not change the intensity of their response to preserve them.
This results in the population colliding with the boundaries, achieving the effect of pushing them outward, hence increasing the region of allowed stances, indicated by the corresponding volume.

Now, if the population is changing stances rather slowly so that temperature can be considered constant, one can observe that pressure and volume relate inversely: when one increases, the other decreases.
This can be interpreted as a correspondence between the social resistance to a change in the boundaries set by norms and the size of the space of possible stances as described by volume.
Given an analogue of Newton's third law (or a variation of it), this case can also be understood as a reduction (increase) of socially allowed stances corresponding to an effort to strengthen (weaken) relevant norms.

As a consequence, we can obtain similar relations between the total population $N$ and the macroscopic variables in this model: direct for $P$ and $V$, and inverse for $T$.
These also have a natural interpretation that a larger population increases the probability of collisions.

The introduction of chemical potential allows the present framework to describe social systems with variable population. 
In this context, the chemical potential $\mu$ characterizes the statistical tendency of a given social group or stance configuration to gain or lose individuals through population exchange with other groups or with the surrounding social environment.

In analogy to equilibrium statistical mechanics, two social groups capable of exchanging individuals tend toward configurations with equal chemical potential. 
This means that differences in chemical potential are expected to generate net population exchange between the corresponding social groups until equilibrium is reached.
This means that larger chemical potentials may be associated with social
configurations that are statistically more effective at sustaining or
attracting population under the imposed macroscopic conditions.

The existence of distinct chemical potentials also suggests the possibility of treating different communities, ideological sectors, organizations or social identities as different social species.
In this picture, equilibrium corresponds not to the absence of dynamics, but
rather to a statistically stationary balance of population exchange among the
different sectors of the social system.

Within this framework, gradients in chemical potential can generate effective social diffusion processes involving population exchange among different social configurations. 
Such processes may be relevant for describing migration between communities, ideological realignments, or changes in social affiliation.

Naturally, these interpretations should be understood at the level of collective statistical behavior rather than individual intentional actions.
The chemical potential therefore constitutes another emergent macroscopic observable, but now associated with population exchange within the effective thermodynamic description adopted here.

\section{Conclusions}\label{Conclusions}
Several limitations of the present framework must be emphasized. 
The model assumes equilibrium, neglects explicit interactions among individuals, and treats the social system through effective macroscopic variables inspired by thermodynamics. 
Therefore, the resulting description is not intended as a fundamental theory of human behavior, but rather as an exploratory phenomenological framework aimed at investigating whether certain collective social phenomena may admit useful statistical descriptions analogous to those found in physics.

With that in mind, the objective of this work is to investigate whether an effective thermodynamic description of a social system can emerge from the statistical treatment of individual stances.
This is carried out analogously to how thermodynamics emerges from a many-particle description of gases in physics.

The description of individual stances as positions and their corresponding mechanics was recovered from a previous work in which a Newtonian approach was used.
Newton's first law is postulated as in physics, but the inertial mass concept is slightly generalized by supposing that it is dependent on position.
Such a mass function characterizes the individual.
The second law is assumed to be unchanged in terms of forces producing a temporal change in momentum defined as $p=m\dot{x}$, leading to equation (\ref{EoM}).
Newton's third law is not asserted in full; while one can conceive actions having reactions in social contexts, we cannot guarantee that the magnitude of the reaction is equal to that of the action.
Possible asymmetries would need to be incorporated.

Departing from that microscopic setting, we assumed a Boltzmann distribution and explored the simple case of no interactions among individuals.
To achieve this, we assumed that energy (at least kinetic energy) can be defined similarly to the way it is done in physics: $E=\frac{p^2}{2m}$.

Following statistical mechanics, the corresponding partition function is introduced and equation (\ref{EOS}) is used to obtain the equation of state (\ref{SEOS}).
For the particular mass function considered in this work, the resulting
equation of state takes the same form as the ordinary ideal gas law.
This relates volume, pressure, temperature, and number of individuals.
Volume is interpreted as the extension of the allowed region of stances about a social question whose boundaries are set by norms; pressure describes the effort of the population to expand the said region by stances reaching its boundaries; and temperature is a measure of the how fast the individuals are changing their stances on the topic under consideration.

These are social variables in the sense that they are macroscopic.
They are meaningful when interpreted statistically for a group of individuals as a whole.
The interpretation of the resulting ideal-gas-like equation of state is
given in those terms and is found to coherently describe the relationships
among these variables at the macroscopic level.
Also, the fact that the detailed parameters characterizing the individual mass functions do not explicitly appear in the final equation of state suggests a form of macroscopic universality within the present idealized framework.

The introduction of chemical potentials also suggests possible extensions toward multispecies social systems, open populations, migration dynamics, and transitions among competing social groups. 
Such generalizations may provide a useful framework for studying collective processes involving population exchange and dynamical reorganization in complex social systems.

Now, some caveats are required regarding the plausibility of observing phenomena describable by the approach presented here.
Note that in physics, one can prepare systems to obtain and even define variables empirically, like, for example, temperature, where one puts two systems in contact and waits for both to reach thermal equilibrium, and transitively extends this procedure to other systems.
Thus, we obtain a relational property that defines the temperature.
In the case of social systems, this is harder to achieve since one cannot just prepare and dispose of societies.
Instead, the exploration of social thermodynamics requires an approach similar to how one studies astronomical objects, observable but not controllable (at least not fully controllable in the case of societies).
One might use variables analogous to $P$, $V$, $T$, or empirical proxies for
them, as collective observables characterizing the macroscopic state of the
system and assess how well the effective framework captures the relations
among such variables.

Lastly, let us consider some directions for future work.
A natural continuation of the present work is the inclusion of interactions among particles, leading to statistical systems analogous to non-ideal gases.
Such extensions may allow the study of collective effects including clustering, polarization, metastability, and phase transitions. 
Another prospective direction is the development of kinetic descriptions based on distribution functions in phase space, potentially leading to transport equations analogous to those of kinetic theory.

\end{document}